# Advances in Shannon-Based Communications and Computations Approaches to Understanding Information Processing in the Brain

James Tee, *Member, IEEE*, Giorgio M. Vitetta, *Senior Member, IEEE*,
and Desmond P. Taylor, *Life Fellow, IEEE*

*Abstract*—This article serves as a supplement to the recently published call for participation in a Research Topic [1] that is timed to commemorate the 75th anniversary of Shannon's pioneering 1948 paper [2]. Here, we include some citations of key and relevant literature, which reflect our opinions/perspectives on the proposed topic, and serve as guidance to potential submissions.

## I. Introduction

In the context of communications and computations, Claude E. Shannon is well-known for at least three things. First, the *source coding theorem* (i.e., noiseless coding theorem), which defines the maximum limit of data compression (e.g., the minimum number of bits required to represent audio music) [2]. Second, the *noisy-channel coding theorem*, which defines the maximum rate at which information can be transmitted almost error-free through a noisy channel (e.g., the maximum number of bits per second that can be transmitted on broadband internet can transmit) [2]. Third, through his MIT master's thesis, the implementation of Boolean algebra (i.e., the AND, OR, NOT and XOR operations) using electric circuits based on relays and switches [3]; this subsequently became the basis of all modern transistor-based computers. Thus, Shannon is the father of both information theory and modern computing. Shannon's key discoveries on communications and computations serve as the foundational basis for understanding all information processing systems, including the brain.

In all modern computers, communications precede and succeed computation. For example, video data (e.g., MPEG-2) must be transmitted to our computers (e.g., via broadband internet) before the data is processed (e.g., video decoding is executed). Even within the computer itself, data must be exchanged between its different devices (e.g., its hard drive and its random access memory) before computations can take place (inside its microprocessor). Once decoded, video information must be further transmitted to the computer screen. Both communications and computations aspects are crucial to information processing. The circumstance is similar in the brain. For example, visual information acquired through eyes must be transmitted from rods and cones contained in the retinas to the visual cortex in the occipital lobe before such visual information can be processed (i.e., computed). From there, the processed information are further transmitted to other locations of the brain (e.g., to the prefrontal cortex) [4]. Despite the importance of both computations and communications, neuroscience research has traditionally focused predominantly on computational aspects, with communications largely being neglected. For example, David Marr's influential hypothesis on the three levels that *"an information-processing device must be understood before one can be said to have understood it completely"* (namely, computational theory, representation & algorithm, and hardware implementation) excluded any communications aspect [5].

Research momentum and advances have, however, begun to shift very recently. For example, it has now been estimated that communications consume 35 times more energy than computations in the human cortex [6]. Furthermore, a recent exhumation of the long-forgotten discovery that single cell organisms (Paramecium aurelia) are capable of Pavlovian conditioning called into question the widely held Hebbian synaptic hypothesis [7]. This Research Topic aims to spotlight research works that take into consideration communications aspects in the brain. Specifically, the goal is to gather recent advances that apply Shannon's key discoveries on communication and computation to better understand neuronal information processing. Another concurrent goal is to complete this Research Topic by/before early 2023 to commemorate the 75th anniversary of Shannon's pioneering 1948 paper, "A Mathematical Theory of Communication" [2].

The scope of this Research Topic covers computations and communications aspects in the brain, based on humans and animal/organism models. On computations, we are open to both school of thoughts, namely, Hebbian synaptic hypothesis (e.g., long-term potentiation) [8], [9] and cell-intrinsic hypothesis (e.g., RNA-based memories and computations) [10]-[17], as applied to areas such as perception, cognition, learning, memory, and decision making. On communications, we

James Tee and Desmond P. Taylor are with the Communications Research Group, Department of Electrical & Computer Engineering, University of Canterbury, Christchurch, New Zealand (email: james.tee@canterbury.ac.nz).

Giorgio M. Vitetta is with the Department of Engineering "Enzo Ferrari", University of Modena and Reggio Emilia, Italy. (email: giorgio.vitetta@unimore.it).

encourage submissions that cover fundamental aspects of neuronal communications, such as quantization [18]-[22], error control coding [23], modulation [4], [6], [24], properties of channel noise [25], synchronization [26], inter-symbol interference mitigation [27] (and, more in general, the role of Kalman filtering in this field [28]-[32]), queueing theory [33], [34], Poisson processes [35], [36], energy requirements (signal-to-noise ratio) [24], [37], and error rate estimations [38]-[40]. In terms of methods, we welcome submissions employing mathematical modeling, computer simulation, data analysis, new hypothesis and theory, new methods (e.g., algorithms), and existing methods from communications systems engineering applied to the brain. In terms of manuscript types, we are interested in original research, methods, review, mini review, hypothesis and theory, perspective, brief research report, and opinion.

## REFERENCES

[1] J. Tee, G. M. Vitetta and D. P. Taylor, "Advances in Shannon-based communications and computations approaches to understanding information processing in the brain," *Front. Comput. Neurosci.*, January 27, 2022. [Online]. Available: https://www.frontiersin.org/research-topics/32120/advances-in-shannon-based-communications-and-computations-approaches-to-understanding-information-pr

[2] C. E. Shannon, "A mathematical theory of communication," *Bell Syst. Tech. J.*, vol. 27, no. 3, pp. 379-423, 623-656, Jul./Oct. 1948. https://doi.org/10.1002/j.1538-7305.1948.tb01338.x

[3] C. E. Shannon, "A symbolic analysis of relay and switching circuits," M.S. thesis, Dept. of Elec. Eng., Mass. Inst. Tech., Cambridge, MA, USA, 1940 [Online]. Available: https://dspace.mit.edu/handle/1721.1/11173

[4] J. Tee and D. P. Taylor, "Is information in the brain represented in continuous or discrete form?" *IEEE Trans. Mol. Biol. Multi-scale Commun.*, vol. 6, no. 3, pp. 199-209, Dec. 2020. https://doi.org/10.1109/TMBMC.2020.3025249

[5] D. Marr, *Vision: A Computational Investigation into the Human Representation and Processing of Visual Information*. New York, NY, USA: W. H. Freeman, 1982.

[6] W. B. Levy and V. G. Calvert, "Communication consumes 35 times more energy than computation in the human cortex, but both costs are needed to predict synapse number," *Proc. Natl. Acad. Sci. U.S.A.*, vol. 118, no. 18, pp. 1-12, Apr. 2021 [Online]. Available: https://doi.org/10.1073/pnas.2008173118

[7] S. J. Gershman, P. E. M. Balbi, C. R. Gallistel and J. Gunawardena, "Reconsidering the evidence for learning in single cells," *eLife*, vol. 10, pp. 1-15, Jan. 2021 [Online]. Available: https://doi.org/10.7554/eLife.61907

[8] M. Mayford, S. A. Siegelbaum and E. R. Kandel, "Synapses and memory storage," *Cold Spring Harb. Perspect. Biol.*, vol. 4, no. 6, pp. 1-18, Apr. 2012 [Online]. Available: https://cshperspectives.cshlp.org/content/4/6/a005751

[9] E. R. Kandel, Y. Dudai and M. R. Mayford, "The molecular and systems biology of memory," *Cell*, vol. 157, no. 1, pp. 163-186, Mar. 2014 [Online]. Available: https://doi.org/10.1016/j.cell.2014.03.001

[10] F. Johansson, D.-A. Jirenhed, A. Rasmussen, R. Zucca and G. Hesslow, "Memory trace and timing mechanism localized to cerebellar Purkinje cells," *Proc. Natl. Acad. Sci. U.S.A.*, vol. 111, no. 41, pp. 14930-14934, Oct. 2014 [Online]. Available: https://doi.org/10.1073/pnas.1415371111

[11] C. R. Gallistel, "The coding question," *Trends Cogn. Sci.*, vol. 21, no. 7, pp. 498-508, Jul. 2017. https://doi.org/10.1016/j.tics.2017.04.012

[12] A. R. Gold and D. L. Glanzman, "The central importance of nuclear mechanisms in the storage of memory," *Biochem. Biophys. Res. Commun.*, vol. 564, pp. 103-113, Jul. 2021. https://doi.org/10.1016/j.bbrc.2021.04.125

[13] C. R. Gallistel, "The physical basis of memory," *Cognition*, vol. 213, pp. 1-6, Aug. 2021. https://doi.org/10.1016/j.cognition.2020.104533

[14] S. Prasada, "The physical basis of conceptual representation – An addendum to Gallistel (2020)," *Cognition*, vol. 214, pp. 1-2, Sept. 2021. https://doi.org/10.1016/j.cognition.2021.104751

[15] W. T. Fitch, "Information and the single cell," *Curr. Opin. Neurobiol.*, vol. 71, pp. 150-157, Dec. 2021. https://doi.org/10.1016/j.conb.2021.10.004

[16] H. Akhlaghpour, "An RNA-based theory of natural universal computation," *J. Theor. Biol.*, vol. 537, pp. 1-19, Mar. 2022. https://doi.org/10.1016/j.jtbi.2021.110984

[17] F. Baluška and M. Levin, "On having no head: Cognition throughout biological systems," *Front. Psychol.*, vol. 7, 902, pp. 1-19, Jun. 2016. https://doi.org/10.3389/fpsyg.2016.00902

[18] R. M. Gray and D. L. Neuhoff, "Quantization," *IEEE Trans. Inf. Theory*, vol. 44, no. 6, pp. 2325-2383, Oct. 1998. https://doi.org/10.1109/18.720541

[19] J. Z. Sun, G. I. Wang, V. K. Goyal and L. R. Varshney, "A framework for Bayesian optimality of psychophysical laws," *J. Math. Psychol.*, vol. 56, no. 6, pp. 495-501, Dec. 2012. https://doi.org/10.1016/j.jmp.2012.08.002

[20] L. R. Varshney and K. R. Varshney, "Decision making with quantized priors leads to discrimination," *Proc. IEEE*, vol. 105, no. 2, pp. 241-255, Feb. 2017. https://doi.org/10.1109/JPROC.2016.2608741

[21] J. Tee and D. P. Taylor, "A quantized representation of probability in the brain." *IEEE Trans. Mol. Biol. Multi-scale Commun.*, vol. 5, no. 1, pp. 19-29, Oct. 2019. https://doi.org/10.1109/TMBMC.2019.2950182

[22] J. Tee and D. P. Taylor, "A quantized representation of intertemporal choice in the brain." *IEEE Trans. Mol. Biol. Multi-scale Commun.*, vol. 7, no. 1, pp. 1-9, Mar. 2021. https://doi.org/10.1109/tmbmc.2020.3025244

[23] J. Tee and D. P. Taylor, "What if memory information is stored inside the neuron, instead of in the synapse?" *arXiv*, pp. 1-8, Jan. 2021. https://doi.org/10.48550/arXiv.2101.09774

[24] T. Berger and W. B. Levy, "A mathematical theory of energy efficient neural computation and communication," *IEEE Trans. Inf. Theory*, vol. 56, no. 2, pp. 852-874, Feb. 2010. https://doi.org/10.1109/TIT.2009.2037089

[25] A. A. Faisal, L. P. J. Selen and D. M. Wolpert, "Noise in the nervous system," *Nat. Rev. Neurosci.*, vol. 9, no. 4, pp. 292-303, Apr. 2008. https://doi.org/10.1038/nrn2258

[26] S. Ghavami, V. Rahmati, F. Lahouti and L. Schwabe, "Neuronal synchronization can control the energy efficiency of inter-spike interval coding," *IEEE Trans. Mol. Biol. Multi-scale Commun.*, vol. 4, no. 4, pp. 221-236, Dec. 2018. https://doi.org/10.1109/TMBMC.2019.2937291

[27] R. E. Lawrence and H. Kaufman, "The Kalman filter for the equalization of a digital communications channel," *IEEE Trans. Commun.*, vol. 19, no. 6, pp. 1137-1141, Dec. 1971. https://doi.org/10.1109/TCOM.1971.1090786

[28] R. E. Kalman, "A new approach to linear filtering and prediction problems," *J. Basic Eng.*, vol. 82, no. 1, pp. 35-45, Mar. 1960. https://doi.org/10.1115/1.3662552

[29] R. C. Wilson and L. H. Finkel, "A neural implementation of the Kalman filter," *Conf. Proc. NIPS*, pp. 2062-2070, Dec. 2009 [Online]. Available: https://proceedings.neurips.cc/paper/2009/file/6d0f846348a856321729a2f36734d1a7-Paper.pdf

[30] S. J. Schiff, "Kalman meets neuron: The emerging interaction of control theory with neuroscience," *Conf. Proc. IEEE. Eng. Med. Biol. Soc.*, pp. 3318-3321, Sept. 2009. https://doi.org/10.1109/IEMBS.2009.5333752

[31] B. Millidge, A. Tschantz, A. K. Seth and C. L. Buckley, "Neural Kalman filtering," *arXiv*, pp. 1-16, Apr. 2021. https://doi.org/10.48550/arXiv.2102.10021

[32] J. Kern, E. Dupraz, A. Aissa-El-Bey, L. R. Varshney and F. Leduc-Primeau, "Optimizing the energy efficiency of unreliable memories for quantized Kalman filtering," *arXiv*, pp. 1-29, Sept. 2021. https://doi.org/10.48550/arXiv.2109.01520

[33] Y. Liu, "Queuing network modeling of elementary mental processes," *Psychol. Rev.*, vol. 103, no. 1., pp. 116-136, Jan. 1996. https://doi.org/10.1037/0033-295x.103.1.116

[34] Y. Liu, "Queuing and network models," in *The Oxford Handbook of Cognitive Engineering*, J. D. Lee and A. Kirlik, Eds. New York, NY, USA: Oxford University Press, 2013, pp. 449-464. [Online]. Available: http://www-personal.umich.edu/~yililiu/QN-Chapter-2013.pdf

[35] D. Heeger, "Poisson model of spike generation," Teaching Handout, Center for Neural Science, New York University, NY, USA, Sept. 2000 [Online]. Available: http://www.cns.nyu.edu/~david/handouts/poisson.pdf

[36] A. I. Weber and J. W. Pillow, "Capturing the dynamical repertoire of single neurons with Generalized Linear Models," *Neural Comput.*, vol. 29, no. 12., pp. 3260-3289, Dec. 2017. https://doi.org/10.1162/neco_a_01021

[37] A. Manwani and C. Koch, "Detecting and estimating signals over noisy and unreliable synapses: Information-theoretic analysis," *Neural Comput.*, vol. 13, no. 1, pp. 1-33, Jan 2001. https://doi.org/10.1162/089976601300014619

[38] B. Maham, "A communication theoretic analysis of synaptic channels under axonal noise," *IEEE Commun. Lett.*, vol. 19, no. 11, pp. 1901-1904, Nov. 2015. https://doi.org/10.1109/LCOMM.2015.2478006

[39] K. Aghababaiyan and B. Maham, "Error probability analysis of neuro-spike communication channel," *IEEE Symposium on Computers and Communications*, pp. 932-937, Jul. 2017. https://doi.ieeecomputersociety.org/10.1109/ISCC.2017.8024645

[40] W. J. Johnston, "Reliable and context-dependent computation in the brain," Ph.D. dissertation, Faculty of the Division of the Biological Sciences and the Pritzker School of Medicine, University of Chicago, Chicago, IL, USA, Dec. 2020 [Online]. Available: https://knowledge.uchicago.edu/record/2720?ln=en

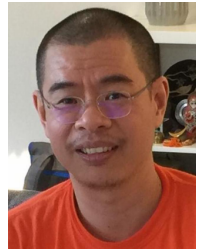

James Tee (Member, IEEE) completed his Ph.D. in Electrical & Electronic Engineering at the University of Canterbury in 2001, where he worked on Turbo Codes under the supervision of Des Taylor. Subsequently, he held various industry and policy positions at Vodafone Group, the World Economic Forum, New Zealand's Ministry of Agriculture & Forestry, and the United Nations. To facilitate his career transitions, he pursued numerous supplementary trainings, including an MBA at the Henley Business School, and an MPhil in Economics (Environmental) at the University of Waikato. In 2012, James began his transition into scientific research at New York University (NYU), during which he completed an MA in Psychology (Cognition & Perception) and a PhD in Experimental Psychology (Neuroeconomics). Afterwards, he worked as an Adjunct Assistant Professor at NYU's Department of Psychology, and a Research Scientist (Cognitive Neuroscience) at Quantized Mind LLC. Since 2017, he is an Adjunct Research Fellow at the University of Canterbury. James is an Eastern medicine physician, with an MS in Acupuncture from Pacific College of Oriental Medicine. In August 2020, he began further training in Substance Abuse Counseling at the New School for Social Research (NSSR). His current research interests in neuroscience focuses on reverse engineering the communications codebook (i.e., signal constellation) of the Purkinje cell neuron. James is also working on Artificial Intelligence (AI) approaches inspired by insights drawn from psychology (cognition, perception, decision-making) and neuroscience.

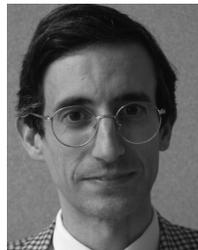

Giorgio M. Vitetta (Senior Member, IEEE) received the Dr.Ing. degree (cum laude) in electronic engineering and the Ph.D. degree from the University of Pisa, Italy, in 1990 and 1994, respectively. He has been holding the position of a full professor of telecommunications at the University of Modena and Reggio Emilia, since 2001. He has coauthored more than 100 papers published on international journals and the proceedings of international conferences. He has coauthored the book Wireless Communications: Algorithmic Techniques (John Wiley, 2013). His main research interests include wireless and wired data communications, localization systems, MIMO radars, and the smart grid. He has served as an Area Editor for the IEEE Transactions on Communications and an Associate Editor for the IEEE Wireless Communications Letters and the IEEE Transactions on Wireless Communications.

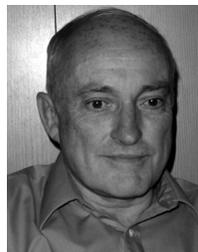

Desmond P. Taylor (Life Fellow, IEEE) received the Ph.D. degree in electrical engineering from McMaster University, Hamilton, ON, Canada, in 1972. From 1972 to 1992, he was with the Communications Research Laboratory and the Department of Electrical Engineering, McMaster University. In 1992, he joined the University of Canterbury, Christchurch, New Zealand, as the Tait Professor of communications. He has authored approximately 250 published papers and holds several patents in spread spectrum and ultra-wideband radio systems. His research is centered on digital wireless communications systems focused on robust, bandwidth-efficient modulation and coding techniques, and the development of iterative algorithms for joint equalization and decoding on fading, and dispersive channels. Secondary interests include problems in synchronization, multiple access, and networking. He is a Fellow of the Royal Society of New Zealand, the Engineering Institute of Canada, and the Institute of Professional Engineers of New Zealand.